\documentclass[useAMS,usenatbib]{mn2e}

\usepackage{comment}
\usepackage{amsmath}
\usepackage{graphicx}
\allowdisplaybreaks[1]

\title[The Hall effect in accretion flows]{The Hall effect in accretion flows}
\author[C. Braiding and M. Wardle]{C. R. Braiding\thanks{E-mail:
catherine.braiding@gmail.com} and M. Wardle\\
Department of Physics \& Astronomy and Research Centre for Astronomy,
Astrophysics \& Astrophotonics\\
Macquarie University, Sydney, NSW 2109, Australia}
\begin{document}

\date{Modified: \today}

\pagerange{\pageref{firstpage}--\pageref{lastpage}} \pubyear{2012}

\maketitle

\label{firstpage}

\begin{abstract}
Magnetic diffusion in accretion flows changes the structure and angular
momentum of the accreting material. We present two power law similarity
solutions for flattened accretion flows in the presence of magnetic diffusion:
a secularly-evolving Keplerian disc and a magnetically-diluted free fall onto
the central object. The influence of Hall diffusion on the solutions is
evident even when this is small compared to ambipolar and Ohmic diffusion, as
the surface density, accretion rate and angular momentum in the flow all
depend upon the product $\eta_H(\mathbf{B}\cdot\Omega)$, and the inclusion of
Hall diffusion may be the solution to the magnetic braking catastrophe of star
formation simulations. 
\end{abstract}


\section{Introduction}

Accretion flows onto protostars are flattened and braked by magnetic fields.
The transportation of angular momentum from the flow to the envelope is
facilitated by the bending of the magnetic field lines and inhibited by the
drift of those lines against the flow\footnote{This is only strictly true for
ambipolar and Ohmic diffusion.}. In magnetohydrodynamical (MHD) simulations of
star formation it has been shown that it is possible to remove all of the
angular momentum from the accretion flow using a weak field, with the
consequence that no accretion disc forms \citep[e.g.][]{pb2007, ml2008}. 
Magnetic diffusion reduces the braking so that a protostellar disc forms
\citep[e.g.][hereafter KK02; Mellon \& Li 2009]{kk2002}; these discs are
precursors to those in which planet formation occurs. 

The amount of magnetic braking affecting an accretion flow depends on the
coupling of the largely neutral medium to the magnetic field, which is
facilitated by collisions between the neutral and charged particles that
transmit the Lorentz force to neutrals. In most simulations the magnetic field
behaviour has been approximated by ideal MHD (where the magnetic field is
frozen into the neutral particles), however this approximation breaks down as
the density in the disc increases. The relative drifts of different charged
species with respect to the neutral particles delineate three conductivity
regimes. \textit{Ohmic (resistive)} and \textit{ambipolar diffusion}, which
behave similarly in a thin disc (to a first order approximation), have been
shown to reduce the effectiveness of magnetic braking \citep{sglc2006,
mim2007, ml2009}. \textit{Hall diffusion} occurs when the degree of coupling
between the different charged species and the neutrals varies, and is expected
to dominate in large regions in accretion flows \citep{ss2002, w2007}. For
certain field configurations Hall diffusion can in principle cause the
magnetic field to be accreted faster than the neutral fluid \citep{wn1999,
bw2011}. 

The role of Hall diffusion in star formation and accretion discs is neither
fully understood nor explored due to the difficulty of performing numerical
simulations \citep[although this is starting to change, e.g.][]{lks2011,
kls2011}, highlighting the need for simpler analytic models that demonstrate
the importance of the Hall effect. In this paper we present two power law
similarity solutions to the MHD equations for an isothermal thin disc with
Hall, ambipolar and Ohmic diffusion: one in which the disc is Keplerian and a
second in which the collapsing material spirals onto the central object
without disc formation. We discuss the implications and the physics
controlling these remarkable solutions. 

\section[]{Formulation}\label{equations}

The magnetohydrodynamic equations for an isothermal system are given by
\begin{equation}
    \frac{\partial\rho}{\partial{t}} + 
    \nabla\cdot(\rho\mathbf{V}) = 0,
    \label{m1}
\end{equation}
\begin{align}
    \rho\frac{\partial\mathbf{V}}{\partial{t}}
    + \rho(\mathbf{V}\cdot\nabla)\mathbf{V} = -\nabla{P} - \rho\nabla\Phi
    + \mathbf{J}\times\mathbf{B},
    \label{rm1}
\end{align}
\begin{align}
    \nabla^2\Phi = {4{\pi}G\rho},
    \label{pot1}
\end{align}
\begin{align}
    \nabla\cdot\mathbf{B}=0,
    \label{am1}
\end{align}
and
\begin{align}
    \frac{\partial{\mathbf{B}}}{\partial{t}} &= \nabla \times (\mathbf{V} 
    \times \mathbf{B}) \nonumber \\ &-\!\nabla\!\times\! 
    \left[\eta\!\left(\nabla\!\times\!\mathbf{B}\right) + 
    \eta_{H}\!\left(\nabla\!\times\!\mathbf{B}\right)\!\times\!\mathbf{\hat{B}} 
    + \eta_{A}\!\left(\nabla\!\times\!\mathbf{B}\right)_{\perp}\!\right]\!,
    \label{in1}
\end{align}
where $\rho$ is the gas density, $\mathbf{V}$ the velocity field, $P$ the
midplane gas pressure, $\Phi$ the gravitational potential (defined as
$\mathbf{g} = -\nabla\Phi$ where $\mathbf{g}$ is the gravitational field), $G$
the gravitational constant, $\mathbf{B}$ the magnetic field and $\mathbf{J}$
the current density defined as $\mathbf{J} = c(\nabla \times \mathbf{B})/4\pi$
where $c$ is the speed of light. The diffusion coefficients for the Ohmic,
Hall and ambipolar terms in the induction equation are $\eta$, $\eta_H$ and
$\eta_A$ respectively. 

We adopt cylindrical coordinates and assume that the disc is axisymmetric and
thin. The velocity field, $\mathbf{V}(r)$, has both radial and azimuthal field
components, the latter giving rise to the specific angular momentum $J = r
V_\phi$, as the material is assumed to settle rapidly to the disc midplane. We
assume that the disc is threaded by an open magnetic field that is symmetric
about the midplane, defining $B_z$ as the vertical field component at the
midplane and $B_{rs}$ and $B_{\phi s}$ as the components at the surface $z =
H$, where $H(r)$ is the half-thickness of the disc. We neglect any mass 
loss due to a disc wind. As the disc is isothermal, the pressure at the
midplane is given by $P = \rho c_s^2$, where $c_s$ is the constant isothermal
sound speed, typically taken to be $0.19$ km s$^{-1}$. 

Equations \ref{m1}--\ref{in1} are vertically-averaged as in \citet[following
KK02]{bw2011} by assuming the disc is thin and integrating over $z$, which
allows us to discard any terms of order $H/r$. The density, radial velocity,
azimuthal velocity and radial gravity are approximated as being constant with
height, as are the quantities $\eta$, $\eta_H/B$ and $\eta_A/B^2$. The radial
and azimuthal magnetic field components are taken to scale linearly with
height, so that their values at the disc surface ($B_{rs}$ and $B_{\phi s}$)
may be used to characterise the field. 

The equations are further simplified by employing monopole expressions for
$B_{rs}$ and $g_r$: 
\begin{align}
    B_{rs} &= \frac{\Psi(r,t)}{2\pi{}r^2} \label{b_rs1}\\
    \text{and }g_r &= -\frac{GM(r,t)}{r^2},å \label{g_rs1}
\end{align}
where $M$ and $\Psi$ are the mass and magnetic flux enclosed within a radius
$r$. These describe the disc sufficiently well that a more complicated
iterative method of calculating the field and gravity is unnecessary
\citep{cck1998}. 

The azimuthal field component is calculated by balancing the torques on the
disc from rotation and magnetic braking, which transports angular momentum
from the disc to the low-density external medium into which the field is
frozen \citep{bm1994}. Assuming that the background angular frequency is small
and that the external Alfv\'en wave speed is constant and parameterised by
$V_\text{A,ext} = c_s/\alpha$, the azimuthal field component is then
\begin{align}
    B_{\phi s}=-\mathrm{min}\Biggl[\frac{\Psi\alpha}{\pi{r^2}c_s}
     &\left[\frac{J}{r}-\frac{\eta_HB_{rs}}{HB}\right]\nonumber\\
     &\left[1+\frac{\Psi\alpha}{\pi{r^2}c_s}\frac{\eta_A}{B^2}
       \frac{B_z}{H}\right]^{-1};\delta{B_z}\Biggr]
\label{b_phisfinal}
\end{align}
\citep{bw2011}. The field is capped as a way of representing the many
magnetohydrodynamical instabilities and processes that might prevent the
azimuthal field component from exceeding the poloidal component such as
internal kinks, turbulence and the magnetorotational instability (MRI). The 
cap $\delta = 1$ corresponds to the typical value of $|B_{\phi s}|/B_z$
obtained when the vertical angular momentum transport is dominated by a
centrifugally-driven disc wind (KK02). 

The vertically-averaged equations are then:
\begin{align}
    &\frac{\partial\Sigma}{\partial{t}} + 
	\frac{1}{r}\frac{\partial}{\partial{r}}(r\Sigma{V_{r}}) = 0,
	\label{mass1}\\ \nonumber\\
    &\frac{\partial{V_{r}}}{\partial{t}} + {V_{r}}\frac{\partial{V_{r}}}{\partial{r}} 
    	= g_r - \frac{c_s^{2}}{\Sigma}\frac{\partial\Sigma}{\partial{r}} 
    	+ \frac{B_{z}B_{rs}}{2\pi\Sigma} + \frac{J^2}{r^3}, \label{rad1}\\
 	\nonumber\\
    &\frac{\partial{J}}{\partial{t}} + V_{r}\frac{\partial{J}}{\partial{r}}
    	= \frac{rB_{z}B_{\phi s}}{2\pi\Sigma},\label{ang1}\\ \nonumber\\
    &\frac{{\Sigma}c_s^2}{2H} 
        = \frac{\pi}{2}G\Sigma^2 + \frac{GM_{c}\Sigma{H}}{4r^3}
        + \frac{1}{8\pi}\left(B_{rs}^{2} + B_{\phi s}^2\right),\label{vert1}
\end{align}
and
\begin{align}
    &\frac{H}{2\pi}\frac{\partial\Psi}{\partial{t}}
        = -rHV_rB_z - \eta{B_{rs}}
        - \frac{r\eta_H}{B}B_zB_{\phi s}
        - \frac{r\eta_A}{B^2}B_{rs}B_z^2; \label{ind1}
\end{align}
$\Sigma$ is the column density, defined as $\Sigma = 2H\rho$, and $M_c$ is the
mass of the central star. These equations are a simplified set of those used
in \citet{bw2011}, removing the $H\partial{B_z}/\partial{r}$ terms that were
used to refine the structure of the disc and shocks in the full similarity
solutions. 
 
The Ohmic and ambipolar diffusion terms scale together, to a zeroth-order
approximation, as they possess a similar dependence upon $B$ and appear in the
induction equation multiplied with the same field component. Ohmic diffusion
is expected to be more important in the inner regions where the density is
high, while ambipolar diffusion dominates in the outer regions \citep{w2007}.
As the field within the disc is effectively vertical, ambipolar and Ohmic
diffusion influence the field drift in the same direction, and only one term 
is required to study the change in disc behaviour introduced by Hall
diffusion. We then describe ambipolar and Ohmic diffusion using the Pedersen 
diffusivity, 
\begin{equation}
    \eta_P = \eta_A + \eta,
\end{equation}
which we parameterise by the constant nondimensional Pedersen diffusion
parameter, $\tilde{\eta}_P$ (described in Appendix A). 

Similarly, we define a nondimensional Hall diffusion parameter
$\tilde{\eta}_H$ such that the Hall coefficient $\eta_H$ scales with the 
surface density and thickness of the thin disc in a similar way to the
Pedersen coefficient (see Appendix A). The direction of Hall diffusion depends
upon the product $\eta_H(\mathbf{B}\cdot\Omega)$, where the sign of $\eta_H$
depends on microphysics within the disc \citep{wn1999}; in our simulations we
vary the sign of $\tilde{\eta}_H$ in order to examine the effect of reversing
the magnetic field. By studying the range of parameters that give stable
accretion flow solutions it may be possible to place constraints on their
physical values in observed accretion systems.

We look for solutions to the equations that take the form of power laws with
respect to a similarity variable $x = r/c_st$ in the limit that $x\to0$.
These may then be thought of as valid in the limits $r\to0$ or $t\to\infty$,
that is, the innermost regions or late stages of the accretion flow. Only two
physical solutions exist to these equations (see Appendix \ref{Assim}): a
secularly-evolving Keplerian disc and a magnetically-diluted near-free fall
collapse. We present these in the following two sections. 

\section{Keplerian disc solution}\label{kep}

The first solution is a Keplerian disc, where the material is supported
against gravity by rotation ($V_\phi = \pm\sqrt{GM/r}$; the
$\mathbf{z}$-axis is defined such that $B_z$ is always positive) and accretion
onto the central mass is slow. The disc is described by the simplified set of
equations:  
\begin{align}
    g_r &+ \frac{J^2}{r^3} = 0,\label{kepJ}\\
    \frac{\partial{J}}{\partial{r}} &= \frac{rB_{\phi s}B_z}{2\pi\Sigma V_r},\\
    HV_r + \frac{\eta_H}{B}B_{\phi s} &+ \frac{\eta_P}{B}B_{rs} = 0,
       \label{kepeqnsid}\\
    \frac{GM_c}{2r^3}H^2 &+ {\pi}G{\Sigma}H - c_s^2 = 0, \label{kepeqnsvhe}\\
    \dot{M} &= \text{constant},\\
    J &= rV_\phi,\label{kepvphi}\\
    B_{rs} &= \frac{4}{3}B_z,\label{kepeqnsbrs}\\
    B_{\phi s} &= -\delta{B_z},\label{kepeqnsbphs}\\
    \text{and }\Psi &= \frac{8}{3}\pi{r^2}B_z;\label{keppsi}
\end{align}
the other terms in Equations \ref{b_rs1}--\ref{ind1}, particularly the
derivatives with respect to time, are negligible in this limit. The induction
equation (\ref{kepeqnsid}) can be written quite simply as $V_r + V_{Br} = 0$,
that is, the inward radial velocity of the fluid is balanced by the drift of
the magnetic field with respect to the gas ($V_{Br}$). Any accretion through
the centrifugally-supported disc is regulated by the outward diffusion of the
magnetic field against the flow; the magnetic diffusion then determines the
column density $\Sigma$. The accretion rate $\dot{M}$ and the direction of
rotation in the disc are determined by the magnetic torque $B_{\phi s}B_z =
-\delta B_z^2$, and are not affected by a global reversal of the magnetic
field. 

Hall diffusion can either add to or act against the Pedersen diffusion in the
disc, as $\eta_H$ can be either positive or negative, while $\eta$ and
$\eta_A$ are both positive. The cap on the azimuthal field, $\delta$, can have
either sign, creating counter rotating solutions when $\delta$ is negative.
Hall diffusion acts in the opposite direction in the counter rotating
solutions (as the Hall term in Equation \ref{kepeqnsid} has the opposite
sign), causing there to be symmetry between the clockwise solutions with
$\eta_H$ positive and large and the anticlockwise solutions with large
negative $\eta_H$. 

\begin{figure}
  \centering
  \includegraphics[width=84mm]{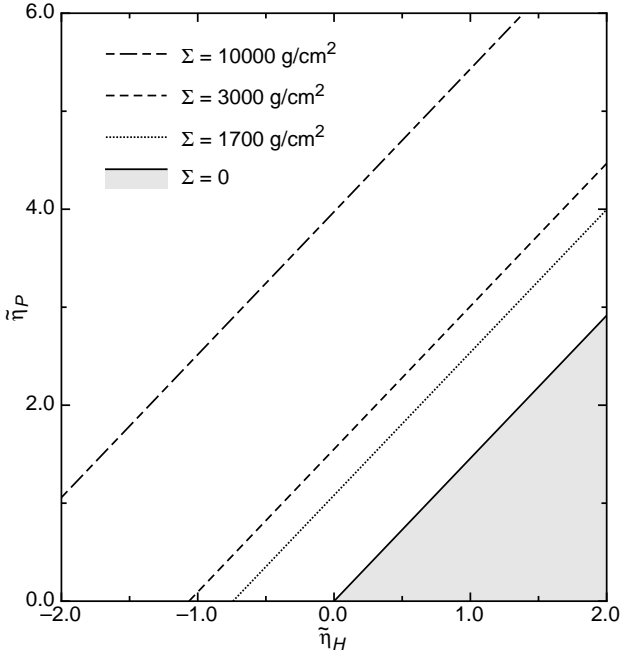}
  \caption{Exploration of the relationship between the Hall and Pedersen
diffusivities $\tilde{\eta}_H$ and $\tilde{\eta}_P$ in the Keplerian disc
solution for various disc surface densities $\Sigma$ at $r = 1$ au and $t =
10^4$ yr, when the azimuthal field cap $\delta = 1$, the sound speed $c_s =
0.19$ km s$^{-1}$ and the accretion rate is $\dot{M} = 10^{-5}$ M$_{\odot}$
yr$^{-1}$ (corresponding to $B_z = 1.15$ G). The solid black line is $\Sigma =
0$, and the shaded region beneath has no solutions describing an accreting
disc rotating in the clockwise direction. The dashed lines correspond to
$\Sigma = 1700$, $3000$ and $10000$ g cm$^{-2}$ respectively; the equivalent
counter rotating solutions are reflected about $\tilde{\eta}_H = 0$.} 
\label{etaKvsetaB}
\end{figure}

The power law solution is given by: 
\begin{align}
    M&=\frac{\dot{m}c_s^3}{G}t,
	&\dot{M} &= \frac{\dot{m}c_s^3}{G},\\ 
    V_r &= -\frac{\dot{m}}{\sigma_1}\left(\frac{rc_s}{t}\right)^{1/2},   
	&\Sigma &= \frac{\sigma_1c_s^{5/2}}{2\pi{G}}\frac{t^{1/2}}{r^{3/2}},\\
    V_\phi &= \pm\sqrt{\frac{\dot{m}c_s^3t}{r}},
        &J &= \pm\sqrt{\dot{m}c_s^3rt}, \\
    B_z &= \frac{(\dot{m}c_s^3)^{3/4}}{\sqrt{2|\delta|G}}\frac{t^{1/4}}{r^{5/4}},
    	&\Psi &=\frac{8\pi(\dot{m}c_s^3)^{3/4}}{3\sqrt{2|\delta|G}}r^{3/4}t^{1/4}, \\
    H &= h_1\frac{r^{3/2}}{\sqrt{c_st}},
\end{align}
where the nondimensional constants characterising the collapse are $\dot{m}$,
the nondimensional infall rate onto the central star, $\delta$, the artificial
cap placed on the azimuthal magnetic field component, and $\tilde{\eta}_H$ and
$\tilde{\eta}_P$, the Hall and Pedersen diffusion coefficients. The sign of
$J$ and $V_\phi$ matches that of $\delta$, as the azimuthal drift of the fluid
against rotation causes the field lines to bend at the disc surface. The
magnitude of the cap on the magnetic braking, not its direction, influences
the vertical field and the surface density. Finally, $\sigma_1$ and $h_1$ are
constants, determined by the expressions 
\begin{align}
    h_1 &= \sqrt{\frac{2}{\dot{m}[1+(f/2\delta)^2]}}\\
    \text{and }\sigma_1 &= \frac{\sqrt{2\dot{m}}f}{2|\delta|\sqrt{(2\delta/f)^2 + 1}}
\label{kepconst}
\end{align}
(from Equations \ref{kepeqnsid} and \ref{kepeqnsvhe}), where $f$ is given by
the equation 
\begin{equation}
    f = \frac{4}{3}\tilde{\eta}_P - \delta\tilde{\eta}_H\sqrt{\frac{25}{9} +
	\delta^2}. \label{f}
\end{equation}
This coefficient must be positive for accretion to occur through the disc ---
the opposite sign corresponds to a solution in which the field is diffusing
inwards against the outwards flow of matter. For the typical value of
$\delta = 1$, the condition $f \ge 1$ becomes 
\begin{equation}
    \tilde{\eta}_P \ge \tilde{\eta}_H\sqrt{\frac{17}{8}},
\label{kep-fineq}
\end{equation}
creating a forbidden region of parameter space, shaded in Fig.\
\ref{etaKvsetaB} and bound by the solid line $f = \Sigma = 0$.  

Fig.\ \ref{etaKvsetaB} illustrates the two-dimensional region of magnetic
diffusion parameter space that gives feasible values of the surface density
$\Sigma$ at $r = 1$ au when $t = 10^4$ years, $c_s = 0.19$ km s$^{-1}$ and
$\delta = 1$ for solutions with an accretion rate of $\dot{M}_c = 10^{-5}$
M$_{\odot}$ yr$^{-1}$ (corresponding to $B_z = 1.15$ G). The dotted line
represents a surface density at 1 au of $\Sigma = 1700$ g cm$^{-2}$; this is
the value from the minimum mass solar nebula model \citep{w1977} in which the
surface density of the solar nebula is estimated by adding sufficient hydrogen
and helium to the solid bodies in the solar system to recover standard
interstellar abundances, and spreading this material smoothly into a disc. The
dashed lines correspond to higher surface densities that are more like those
expected to occur in protostellar discs. 

The radial scaling of the surface density ($\Sigma \propto r^{-3/2}$) in this
solution is that expected from the minimum mass solar nebular \citep{w1977}
and other simulations of protostellar discs \citep[e.g.][]{t1999, vb2009}; and
the magnetic field scaling also matches that from theory, particularly of
discs that support disc winds \citep[$B_z \propto r^{-5/4}$;][]{bp1982}. The 
azimuthal field component blows up with respect to the other field components
in this small $x$ limit because the azimuthal magnetic field drift speed is
slow compared to the Keplerian speed. The model adopted for the vertical
angular momentum transport is unable to properly account for the effects of
magnetic braking in the small $x$ limit, so $B_{\phi s}$ must be capped. If a
different scaling for $\eta_H$ were adopted then Hall diffusion could act to
limit $B_{\phi s}$ such that the cap becomes unnecessary; this should be
examined in future disc studies. 

As is clear from Fig.\ \ref{etaKvsetaB}, there is no solution describing a
disc with purely Hall diffusion where the Hall diffusion parameter is
positive, as the positive Hall parameter acts to restrain the effects of
ambipolar and Ohmic diffusion when the disc is rotating in the clockwise
direction. However, when the Hall diffusion parameter is negative
(corresponding to a reversal of the magnetic field) it acts in the same radial
direction as the Pedersen diffusion, enhancing the radial diffusion of the
magnetic field. When the disc is counter rotating, the opposite holds true, so
that the solutions in Fig.\ \ref{etaKvsetaB} are reflected across the vertical
line $\tilde{\eta}_H = 0$. In the case of pure ambipolar diffusion the field
moves inward slower than the neutral particles and the solution reduces to the
fiducial solution of KK02 (their equations 51--57). 

\begin{figure}
  \centering
  \includegraphics[width=82mm]{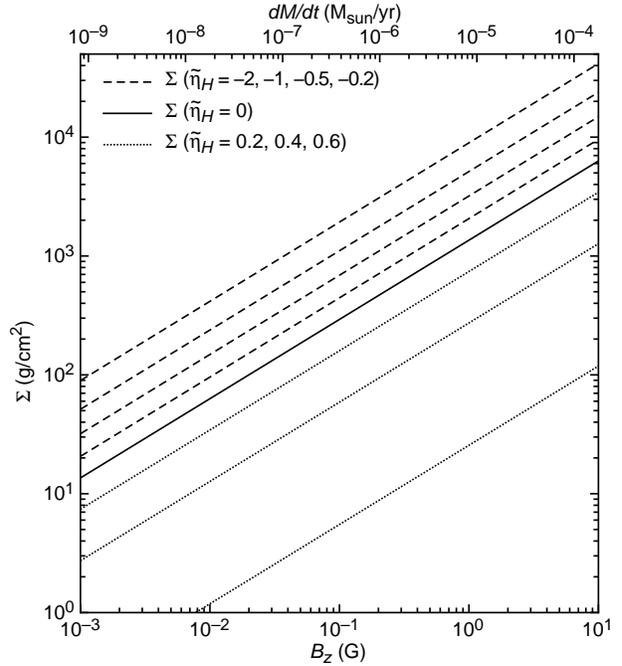}
  \caption{Disc surface density $\Sigma$ against the vertical magnetic field
component $B_z$ and central mass accretion rate $\dot{M}$ for different values
of the Hall diffusion parameter $\tilde{\eta}_H$ for a Keplerian disc rotating
in the clockwise direction with Pedersen parameter $\tilde{\eta}_P = \delta =
1.0$ and $c = 0.19$ km s$^{-1}$, at $r = 1$ au and $t = 10000$ yr. The solid
line corresponds to $\tilde{\eta}_H = 0$; the dashed lines above are
$\tilde{\eta}_H = -0.2, -0.5, -1.0$ and $-2$ respectively; and the dotted
lower lines are $\tilde{\eta}_H = 0.2, 0.4$ and $0.6$.} 
\label{SigmavsB}
\end{figure}

Fig.\ \ref{SigmavsB} plots $\Sigma$ against $B_z$ and $\dot{M}$, showing the
influence of Hall diffusion on $\Sigma$ at 1 au of a disc with $\tilde{\eta}_P
= \delta = 1$. The surface density increases with the negative Hall parameter
to create heavier discs, while a positive value of $\tilde{\eta}_H$ causes
$\Sigma$ to drop off dramatically as $\tilde{\eta}_H$ approaches the limit of
$f \ge 0$. Once more the opposite behaviour holds true when the disc is
counter rotating.  

Equations \ref{kepJ}--\ref{kepvphi} may be written in a more physical form by
writing the variables as functions of $B_z$, $r$, $t$, the diffusion parameter
$f$ and the azimuthal field cap. The magnetic field components are then given
by Equations \ref{kepeqnsbrs}--\ref{keppsi}, while the other variables are: 
\begin{align}
    \dot{M} &= \left(\frac{(2\delta)^2B_z^4}{Gt}r^5\right)^{1/3},\\
    M &= \left(\frac{(2\delta t)^2B_z^4}{G}r^5\right)^{1/3},\\
    \Sigma &= \frac{c_sf}{2\pi|\delta|}\sqrt{\frac{M}{2G[(2\delta/f)^2+1]}}
	r^{-3/2}\,,\label{b_zsetsig}\\
    H &= \sqrt{\frac{2}{G[1+(f/2\delta)^2]M}}c_sr^{3/2}\,, \label{b_zseth}\\
    J &= (2\delta{G}B_z^2r^4t)^{1/3},\\
    V_{\phi} &= (2\delta{G}B_z^2rt)^{1/3}\\
    \text{and }V_r &= -\frac{\delta{B_z^2r}}{\pi{V_\phi}\Sigma}.\label{b_zset}
\end{align}
This model description shows more clearly that the direction of rotation in
the disc is tied to the direction of magnetic braking and emphasizes the build
up of the central mass occurs through accretion against magnetic field
diffusion. 

Alternatively, the disc variables can be thought of as functions of $M$ and
$\dot{M}$, by treating $t$ as the ratio of $M$ to $\dot{M}$ rather than the
age of the system. The solution can then be regarded as a steady state
Keplerian disc, where the variables are defined by 
\begin{align}
    B_z &= \left(\frac{GM\dot{M}^2}{(2\delta)^2r^5}\right)^{1/4},\\
    J &= \sqrt{GMr}\,,\\
    V_{\phi} &= V_K = \sqrt{\frac{GM}{r}}\,\\
    \text{and }V_r &= -\frac{\dot{M}}{2\pi\Sigma{r}}\,;
\label{mset}
\end{align}
$\Sigma$ and $H$ are unchanged from Equations \ref{b_zsetsig} and
\ref{b_zseth}, and the field components are as in Equations
\ref{kepeqnsbrs}--\ref{keppsi}. 

The amount of Hall diffusion in the flow helps determine how much matter is
accreted onto the central object compared with that remaining in the disc; 
decreasing the Hall parameter $\tilde{\eta}_H$ in comparison to a constant
$\tilde{\eta}_P$ causes the surface density become large as the radial field
diffusion slows the rate of infall. The ambipolar diffusion-only fiducial
solution of KK02 has an accretion rate of $\dot{M}_c = 7.66 \times 10^{-6}$
M$_\odot$ yr$^{-1}$, and the surface density at 1 au is $1310$ g cm$^{-2}$.
Adding a positive value of $\tilde{\eta}_H$ to this solution causes the
surface density to decrease, which is problematic if one expects to form large
planets in a protostellar disc, however, if the Hall parameter were negative
then the surface density would be raised to a more realistic value. This again
is only true for solutions rotating in the clockwise direction; however, as
the sign of $\eta_H$ depends on microphysics in the disc, the general point
that planetary system architecture depends on the product $\eta_H
(\mathbf{B}\cdot\Omega)$ stands. 

\section{Free Fall Solution}\label{ff}

The second similarity solution describes the infall when the magnetic braking
is very efficient at removing angular momentum from the flow. In this case
there is very little angular momentum remaining and the reduced centrifugal
support inhibits disc formation, so that the collapsing flow becomes a
supersonic magnetically-diluted free fall onto the central protostellar mass.
This solution is representative of the magnetic braking catastrophe that
affects many numerical simulations of gravitational collapse. The collapse is
described by the equations: 
\begin{align}
    V_r\frac{\partial{V_r}}{\partial{r}} &= g_r,\label{ff-eqnsvr}\\
    \frac{\partial{J}}{\partial{r}}
      &= \frac{rB_{\phi s}B_z}{2\pi\Sigma{V_r}},\\
    V_rH &+ \frac{\eta_H}{B}B_{\phi s} + \frac{\eta_P}{B^2}B_{rs}B_z = 0,\\
    \frac{GM}{2r^3}H^2 &+ \frac{(B_{rs}^2 + B_{\phi s}^2)}{4\pi\Sigma}\,H
      -c_s^2 = 0,\\
    \dot{M} &= \text{constant}\\
    J &= rV_\phi,\label{ff-eqnsJ}\\
    B_{rs} &= B_z\label{ff-eqnsbrs}\\
    \Psi &= {2\pi} r^2B_z\label{ff-eqnspsi}\\
    \text{and }B_{\phi s} &= -\text{min}\left[\frac{-\eta_H}{\eta_P}\,B\,;\,
      \delta B_z\right];\label{ff-eqnsbphis} 
\end{align}
as in the Keplerian disc solution the induction equation takes the simplified
form $V_r + V_{Br} = 0$, however here the magnetic diffusion is assumed to be 
large and dynamically important. In this similarity solution any rotation
remaining in the flow is that induced by the magnetic ``braking'', which can
cause rotation by Hall diffusion of the field lines tied to the electrons in
the azimuthal direction, which creates a rotational torque on the neutrals and
grains as they fall inward rapidly. 

The dimensional form of this similarity solution is given by the complete set
of fluid variables: 
\begin{align}
    M &= \frac{\dot{m}c_s^3}{G}t,
	&\dot{M} &= \frac{\dot{m}c_s^3}{G}\label{ff-mmdot}\\ 
    V_r &= -\sqrt{\frac{2\dot{m}c_s^3t}{r}},
	&\Sigma &= \frac{1}{\pi G}\sqrt{\frac{\dot{m}c_s^3}{2^3rt}}, \label{ff-vsig}\\
    V_\phi &= j_{1,2}c_s,
	&J &= j_{1,2}c_sr, \label{ff-vj}\\
    B_z &= \frac{c_s^2b_{z1,2}}{G^{1/2}}r^{-1}, 
        &\Psi &= \frac{2\pi c_s^2b_{z1,2}}{G^{1/2}}r,\label{ff-psibz}\\
    B_{rs} &= B_z,
        &H &= h_{1,2}\,\frac{r^{3/2}}{\sqrt{c_st}}, \label{ff-h}
\end{align}
and
\begin{equation}
    B_{\phi s} = -\text{min}\left[\frac{-\tilde{\eta}_H}{\tilde{\eta}_P}B\,;\,
		\delta B_z\right]. \label{ff-bphi}
\end{equation}
The sign of the Hall diffusion coefficient and the cap on $B_{\phi s}$ must be
taken into account when calculating the minimum of Equation \ref{ff-bphi}, as
this should be an absolute minimum. We define the $\mathbf{\hat{z}}$ direction
such that $B_z$ is always positive, and as the radial velocity is negative
then conservation of angular momentum (Equation \ref{ff-eqnsJ}) requires
$B_{\phi s}$ and $J$ to have the opposite sign. Any twisting of the field by
Hall diffusion is balanced by spinning the fluid in the opposite direction. 

Given that there are two possible values for $B_{\phi s}$, depending upon the
chosen values of the parameters $\tilde{\eta}_{H,P}$ and $\delta$, there are
two possible coefficients for $B_z$, $J$ and $H$. These are referenced by the
subscripts 1 and 2 depending on which term in Equation \ref{ff-bphi} is
smaller. The first set of coefficients occur when 
\begin{equation}
    \sqrt{\frac{2\tilde{\eta}_H^2}{\tilde{\eta}_P^2 - \tilde{\eta}_H^2}} 
	< |\delta|,
\label{b_phisineq}
\end{equation}
so that the azimuthal field is given by 
\begin{equation}
    B_{\phi s} = \frac{\sqrt{2}\tilde{\eta}_HB_z}{\sqrt{\tilde{\eta}_P^2 -
	\tilde{\eta}_H^2}}. \label{ff-b_phis1}
\end{equation}
The vertical field coefficient $b_{z1}$ is then the single real root of the
polynomial 
\begin{equation}
    b_{z1}^8 - \frac{\dot{m}^2}{2\tilde{\eta}_P^2f_1} \,b_{z1}^6 - 
      \frac{\dot{m}^6}{4\tilde{\eta}_P^4f_1^4} = 0, 
    \label{1-bz1}
\end{equation}
where $f_1$ is a constant defined by the magnetic diffusion parameters:
\begin{equation}
    f_1 = \frac{\tilde{\eta}_P^2 + \tilde{\eta}_H^2}
	{\tilde{\eta}_P^2 - \tilde{\eta}_H^2}.
    \label{1-f}
\end{equation}
The coefficients for $j$ and $h$ are given by
\begin{equation}
    j_1 = -\frac{b_{z1}^2}{\dot{m}} \,\frac{\sqrt{2}\,\tilde{\eta}_H}
	{\sqrt{\tilde{\eta}_P^2 - \tilde{\eta}_H^2}}
    \label{1-j1}
\end{equation}
and
\begin{equation}
    h_1 = \frac{f_1b_{z1}^2}{\sqrt{2\dot{m}^3}}
      \left[-1+\sqrt{1-\frac{4\dot{m}^2}{f_1^2b_{z1}^4}}\,\right].
    \label{1-h1}
\end{equation}
When there is only ambipolar diffusion, this solution reduces to the
nondimensional asymptotic behaviour of the ``strong braking'' collapse of KK02
(their equations 66--71). In their solution $f_1 = 1$ and $j_1 \approx 0$, so
that the angular momentum and the azimuthal field component are reduced to a
small plateau value as the strong magnetic braking removes almost all of the
angular momentum early in the collapse (see their fig.\ 10). Their similarity
solutions tended towards this behaviour when the magnetic braking parameter
$\alpha$ was large, even though it does not appear in the coefficients of
the power law solutions. 

This set of coefficients corresponds to the curves in Fig.\ 3, which plots
$\tilde{\eta}_P$ against $\tilde{\eta}_H$ for cores with varied magnetic
fields and $\dot{M} = 10^{-5}$ M$_\odot$ yr$^{-1}$ at 1 au and $t = 10^4$ yr.
The direction of rotation changes with the sign of $\tilde{\eta}_H$, with the
solid lines denoting negative rotation and the dotted lines positive. As the
field increases the size of the area enclosed by the curve decreases, and the
solutions tend towards the asymptote $\tilde{\eta}_P = |\tilde{\eta}_H\!|$ as
both the diffusivities tend to zero. The $\tilde{\eta}_P$-intercept occurs at 
\begin{equation}
    \tilde{\eta}_P = \frac{\dot{m}}{2b_{z1}}
	\left[1 + \sqrt{1 + \frac{4\dot{m}^2}{b_{z1}^4}}\right]^{1/2},
\label{eta_a}
\end{equation}
corresponding to the value of $\tilde{\eta}_P$ needed to give the same
magnetic field strength in KK02. The solid line intercepting the curves is
$\tilde{\eta}_P = |\tilde{\eta}_H| \sqrt{2+\delta^2}/\delta$, which marks the
boundary between the coefficient sets and in the limit of large $|\delta|$
also tends towards $\tilde{\eta}_P = |\tilde{\eta}_H|$. 
\begin{figure}
  \centering
  \includegraphics[width=84mm]{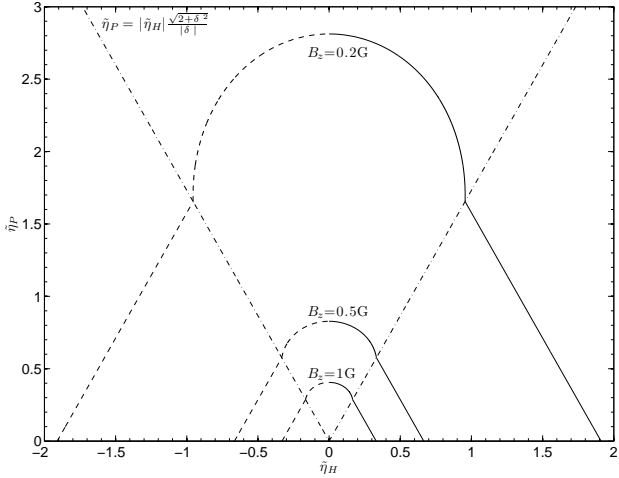}
  \vspace{-5mm}
  \caption{The relationship between the Hall and Pedersen diffusion parameters
$\tilde{\eta}_H$ and $\tilde{\eta}_P$ in the free fall solution, for various
midplane magnetic field strengths $B_z$ at $r = 1$ au and $t=10^4$ yr, where
the azimuthal field cap $\delta=1$ and the sound speed $c_s = 0.19$ km
s$^{-1}$. We take the central mass accretion rate to be $\dot{M} = 10^{-5}$
M$_{\odot}$ yr$^{-1}$ (corresponding to a surface density at $r = 1$ au of
$\Sigma = 5.03$ g cm$^{-2}$). The solid curves correspond to negative
rotation, the dashed ones positive, and the boundary between the two values of
$B_{\phi s}$ is the dot-dashed line $\tilde{\eta}_P = \tilde{\eta}_H
\sqrt{2+\delta^2}/\delta$.} \label{etakvsetab-bz2} 
\end{figure}

The second set of coefficients apply when the inequality
\begin{equation}
    \sqrt{\frac{2\tilde{\eta}_H^2}
	{\tilde{\eta}_P^2 - \tilde{\eta}_H^2}} > |\delta|
\label{b_phisineq2}
\end{equation}
is satisfied, so that $B_{\phi s}$ takes on the second value in Equation
\ref{ff-bphi}, $B_{\phi s} = -\delta B_z$. The coefficient $b_{z2}$ is given
by the single positive real root of the equation 
\begin{equation}
    b_{z2}^8 - \frac{\dot{m}^2(1+\delta^2)}{2f_2^2}\,b_{z2}^6 
	- \frac{\dot{m}^6}{4f_2^4} = 0
    \label{2-bz2}
\end{equation}
where $f_2$ is
\begin{equation}
   f_2 = \tilde{\eta}_P - \tilde{\eta}_H\delta\sqrt{2+\delta^2}.
   \label{2-f}
\end{equation}
This then gives the other coefficients as
\begin{equation}
    j_2 = -\frac{\delta{b_{z2}^2}}{\dot{m}}
    \label{2-j2}
\end{equation}
and 
\begin{equation}
    h_2 = \frac{(1+\delta^2)b_{z2}^2}{\sqrt{2\dot{m}^3}}
	\left[-1+\sqrt{1+\frac{4\dot{m}^2}{(1+\delta^2)^2b_{z2}^4}}\right].
    \label{1-h_2}
\end{equation}
This solution was not explored in KK02, as without Hall diffusion the left
term in Equation \ref{ff-bphi} is zero. 

The second set of coefficients corresponds to the straight lines outside of
the curves in Fig.\ 3, again with negative rotation drawn as solid lines and
positive as dashed. The slope of the line (and the direction of rotation) are
both dependant upon the sign of $\delta$, i.e.\ the cap on $B_{\phi s}$; in
this solution the cap may have either sign, corresponding to the direction in
which the field was azimuthally bending from the vertical before the cap was
attained. Under a global reversal of the field the infalling matter spins in
the opposite direction. 

The equations may also be rearranged to give the variables as functions of the
surface density and magnetic field: 
\begin{align}
    V_r &= -\sqrt{\frac{2GM}{r}}\,,\\
    M &= -2{\pi}r{\Sigma}V_rt\,\\
    \dot{M} &= -2{\pi}r{\Sigma}V_r\\
    \text{and }J &= \frac{B_{\phi s}B_z}{2{\pi}{\Sigma}V_r} \,r^2,
\label{ff-sigmaeqns}
\end{align}
where the field components are defined in Equations
\ref{ff-eqnsbrs}--\ref{ff-eqnsbphis} and the scale height of the disc is given
by the equations: 
\begin{align}
    &H_1 = \frac{f_1B_z^2r^{7/2}t}{\sqrt{2GM^3}}
      \left[-1+\sqrt{1-\frac{4c_s^2M^2}{f_1^2B_z^4r^4t^2}}\, \right] 
      \text{ or}\qquad\quad\\
    &H_2 = \frac{(1+\delta^2)B_z^2r^{7/2}t}{\sqrt{2GM^3}}
      \!\left[\!-1+\!\sqrt{1-\frac{4c_s^2M^2}{(1+\delta^2)^2B_z^4r^4t^2}} \,\right]
    \label{2-H2}
\end{align}
depending on which value of $B_{\phi s}$ is adopted.

This similarity solution exemplifies the magnetic braking catastrophe that
occurs in numerical simulations of star formation when the magnetic braking of
a collapsing core is so strong that all angular momentum is removed from the
gas and it is impossible to form a rotationally-supported disc
\citep[e.g.][]{ml2008, ml2009}. It is clear from Equations \ref{ff-vj}b, 
\ref{ff-bphi} and \ref{1-j1} that when there is no Hall diffusion the magnetic
braking causes $B_{\phi s} = J = 0$, which would prevent Keplerian disc
formation.  

The introduction of Hall diffusion to the similarity solution can cause
additional twisting of the field lines and magnetic braking, or it can cause a
reduction in magnetic braking by twisting the magnetic field lines in the
opposite direction, in effect spinning up the collapse. The direction of the
Hall diffusion depends upon the orientation of the field with respect to the
axis of rotation, and it is obvious that this directionality has an important
effect on the magnetic braking catastrophe. The linear scaling of the angular
momentum with radius (Equation \ref{ff-vj}) suggests that the point mass at
the origin has no angular momentum, however, Hall diffusion will likely ensure
that the angular momentum shall drop to a plateau value similar to that in the
ambipolar diffusion-only collapse of KK02's fig.\ $10$ in full simulations of
such accretion flows.

\section{Discussion}

The two similarity solutions represent both sides of the magnetic braking
catastrophe of star formation. In the first solution, the magnetic braking is
limited and a rotationally-supported disc similar to those in the simulations
of \cite{mim2010} forms, while in the second no disc forms as catastrophic
magnetic braking removes almost all of the angular momentum as in the
simulations of \citet{ml2009}. The magnetic diffusion, in particular Hall
diffusion, is clearly important in determining whether a disc forms, its size
and rotational behaviour and the density profile within the disc.

This can have wide implications on planet formation, for the size and surface
density of the disc  will then depend upon the product $\eta_H (\mathbf{B}
\cdot \Omega)$, where the sign of $\eta_H$ depends upon the abundances of
charged grains and electrons relative to the neutrals in the disc
\citep{wn1999}. This means that the outcome of planet formation in two
otherwise identical discs can be vastly different if they are initially
rotating in opposite directions. Our results in \S\ref{kep} show that one does
not require a large amount of Hall diffusion for this asymmetry to be
observable in the surface density of a protostellar disc. 

Our second solution shows a similar asymmetry caused by Hall diffusion, as all
of the initial angular momentum has been removed from the collapse by magnetic
braking and the only rotation of the flow is that induced by Hall diffusion.
The direction of rotation corresponds directly with the sign of the Hall
parameter, however the angular momentum remains dynamically unimportant. This
solution ought to apply in simulations of accretion flows where the magnetic
braking parameter $\alpha$ and the cap on the azimuthal field $\delta$ are
large, as in the ``strong braking'' star formation similarity solution of
KK02. 

The prescription for the magnetic braking used in this simple model is limited
by our use of a cap placed on $B_{\phi s}$ to account for missing physics such
as non-axisymmetric effects \citep[e.g.\ the magnetorotational
instability;][]{ss2002}. The azimuthal field takes the value of this cap in 
our Keplerian disc solution, and also in the free fall solution when the Hall
diffusivity is large, however the value $\delta = 1$ adopted in our plots is
that expected from a disc with a disc wind. A more complete description of the
angular momentum transport would include the calculation of a disc wind
\citep{bp1982}, as these are observed in simulations of protostellar and other
accretion discs \citep[e.g.][]{mim2007, ml2009}. 

The similarity solutions of \citet{bw2011} all demonstrated Keplerian disc
formation in the inner region of the protostellar collapse of a molecular
cloud core, matching onto the power law solution described in \S\ref{kep} at
the inner boundary. Enforcing disc formation, Hall diffusion was shown to
affect the mass of the protostar and accretion disc, as well as the size and
structure of the disc. The free fall solution has yet to be adopted as a
boundary condition in the full model, but its applicability to collapsing
cores must be studied in future to show the importance of Hall diffusion in
spinning up the flow, facilitating disc formation in order to solve the
magnetic braking catastrophe. 

\section{Conclusions}

We have presented two distinct power law similarity solutions to the MHD
equations describing a flattened accretion flow with magnetic diffusion. The
first of these represented a rotationally-supported disc through which gas is
slowly accreted while the magnetic field diffuses outwards against the flow.
The second was a free fall collapse onto the star in which almost all of the
angular momentum has been removed from the fluid and no rotationally-supported
disc may form. These have been used to show that even a small amount of Hall
diffusion can dramatically change the dynamics of gravitational collapse. 

\section*{Acknowledgments}

This work was supported in part by the Australian Research Council grant
DP0881066.

\appendix
\section[]{self-similar equations}\label{Assim}
Gravitational collapse occurs over many orders of magnitude in density, so
that the point mass has negligible dimensions in comparison with the accretion
flow. The self-similarity of the wave of infall, which propagates outwards at
the speed of sound, arises because of the lack of characteristic time and
length scales. Then the only dimensional quantities are the gravitational
constant $G$, the isothermal sound speed $c_s$, the local radius $r$ and the
instantaneous time $t$, so that, except for scaling factors, all quantities
must be functions of the similarity variable $x$, defined as: 
\begin{equation}
    x = \frac{r}{c_st}.
  \label{x}
\end{equation}
We then define a set of nondimensional fluid variables that depend only upon $x$:
\begin{align}
    \Sigma(r,t) &= \left(\frac{c_s}{2\pi{Gt}}\right)\sigma(x),&
       g_r(r,t) &= \left(\frac{c_s}{t}\right)g(x),\\
    V_r(r,t) &= c_su(u),&
      H(r,t) &= c_sth(x),\\
    V_\phi(r,t) &= c_sv(x),&
         J(r,t) &= c_s^2tj(x),\\
    M(r,t) &= \left(\frac{c_s^3t}{G}\right)m(x),&
    \dot{M}(r,t)&= \left(\frac{c_s^3}{G}\right)\dot{m}(x),\\ 
    \mathbf{B}(r,t) &= \left(\frac{c_s}{G^{1/2}t}\right)\mathbf{b}(x),&
          \Psi(r,t) &= \left(\frac{2\pi{c_s^3t}}{G^{1/2}}\right)\psi(x).
  \label{ssiming}
\end{align}

In the thin disc approximation used here, the diffusivities are assumed to be
constant with height within the disc, and the Ohmic and ambipolar diffusion 
terms scale together to a zeroth-order approximation. They are combined into
the Pedersen diffusivity, which scales within the disc as 
\begin{equation}
\eta_P = \tilde{\eta}_P \frac{c^2tb^2h^{3/2}}{\sigma^{3/2}};
\label{etap}
\end{equation}
the self-similarity of the diffusivity depends upon the implicit relationship
$\rho \propto \rho_i$, which holds true across a large range of densities
\citep{kn2000}. We adopt a similar scaling for the Hall diffusion coefficient
$\eta_H$ as a matter of pragmatism: 
\begin{equation}
\eta_H = \tilde{\eta}_H \frac{c^2tb^2h^{3/2}}{\sigma^{3/2}},
\label{ssetah}
\end{equation}
where $\tilde{\eta}_H$ is the nondimensional Hall diffusion parameter used to
characterise the solutions, the ratio of the ambipolar to Hall terms becomes
the most important factor in describing the magnetic behaviour of the
solutions.  

For convenience we define $w\equiv{}x-u$ and then use the similarity variables
to write Equations \ref{mass1}--\ref{ind1} as: 
\begin{align}
   &\frac{dm}{dx} = x\sigma, \label{sscm}\\
   (1-w^2)\frac{1}{\sigma}\frac{d\sigma}{dx} &= g + \frac{b_{r,s}b_z}{\sigma} 
	+ \frac{j^2}{x^3} + \frac{w^2}{x}, \label{sscrm}\\
   \frac{dj}{dx} &= \frac{1}{w}\biggl(j 
   	- \frac{xb_zb_{\phi,s}}{\sigma}\biggr),\label{ssam}\\
   \frac{\sigma{\dot{m}}}{x^3}h^2 + \bigl(b_{r,s}^2 &+ b_{\phi,s}^2 + \sigma^{2}\bigr)h 
	- 2\sigma = 0, \label{ssvhe}\\
   \psi - xwb_z + \bigl(\tilde{\eta}_Hb_{\phi,s}b 
	&+ \tilde{\eta}_Pb_{r,s}b_z\bigr)xb_zh^{1/2}\sigma^{-3/2} = 0, \label{ssin} \\
   \text{and }&\frac{d\psi}{dx} = xb_z. \label{sspsi}
\end{align}
These are augmented by the supplementary definitions:
\begin{align}
   m &= xw\sigma, \\
   \dot{m} &= -xu\sigma, \\
   \text{and }g &= -\frac{m}{x^2}, \label{ssextra}
\end{align}
while the field components are given by
\begin{align}
   b_{r,s} &= \frac{\psi}{x^2},\\
   b_{\phi,s} = -\mathrm{min}\Biggl[
	\frac{2\alpha\psi}{x^2}&\left[\frac{j}{x} 
	-\frac{\tilde{\eta}_Hh^{1/2}b}{\sigma^{3/2}}\left(b_{r,s} 
	-h\frac{db_z}{dx}\right)\right]\nonumber\\
	&\left[1 + \frac{2\alpha\tilde{\eta}_Ph^{1/2}\psi{b_z}}{x^2\sigma^{3/2}}
        \right]^{-1};\delta{b_z}\Biggr]. 
\label{ssb_phis}
\end{align}

We find the similarity solutions by assuming that the fluid variables take the
form of power law relations in $x$: 
\begin{align}	
\sigma &= \sigma_1x^{-p} \\
b_z &= b_{z1}x^{-q} \\
j &= j_1x^{-r} \label{pls}
\end{align}
where $p$, $q$, and $r$ are real numbers. We substitute these into the
self-similar equations, and by taking the limit as $x\to0$ we find the
dominant terms and solve for the exponents and coefficients of the power law 
relations. There are only two nontrivial solutions \citep[derived
in][]{b2011}: a Keplerian disc with $p = 3/2$, $q = 5/4$ and $r = -1/2$; and
the near-free fall collapse with $p = 1/2$, $q = 1$ and $r = -1$. 

\subsection{Keplerian disc solution}
The first solution is the Keplerian disc presented in \S\ref{kep}, where the
mass is supported against gravity by its angular momentum. The similarity 
solution in nondimensional form is: 
\begin{align}
    m &= \dot{m},
	&\sigma &= \sigma_1x^{-3/2}, \\
    u &= -\frac{\dot{m}}{\sigma_1}x^{1/2},
	&j &= \pm\sqrt{\dot{m}x}, \\
    \psi &= \frac{4}{3}x^2b_z, 
	&b_z &= \frac{\dot{m}^{3/4}}{\sqrt{2|\delta|}}x^{-5/4}, \\
    b_{r,s} &= \frac{4}{3}b_z,
	&b_{\phi,s} &= -\delta{b_z}, \\
    h &= h_1x^{3/2},
    \label{ssimsol1}
\end{align}
where the constants are defined as in \S\ref{kep}. In the ambipolar diffusion
limit these reduce to equations 51--57 of KK02. 

\subsection{Free fall solution}
The second solution is the set of relations that describe a free fall onto the
central mass dominated by magnetic diffusion:
\begin{align}
    m &= \dot{m}, 
	&\sigma &= \sqrt{\frac{\dot{m}}{2x}}, \\
    u &= -\sqrt{\frac{2\dot{m}}{x}}, 
	&j &= j_{1,2}x, \\
    \psi &= b_zx, 
	&b_z &= b_{z1,2}x^{-1}, \\
    b_{r,s} &= b_z,
    	&b_{\phi,s} &= -\text{min}\left[\frac{\tilde{\eta}_H}{\tilde{\eta}_P}b\,;\,
	{\delta}b_z\right]\!,\label{ssimsol2} \\
    h &= h_{1,2}x^{3/2}.
\end{align}
Given that there are two possible values for $b_{\phi,s}$, depending on the
chosen values of the parameters $\tilde{\eta}_{H,A}$ and $\delta$, there are
two sets of coefficients for $j$, $b_z$ and $h$. These are defined in
\S\ref{ff} and referenced by the subscripts 1 and 2 depending on which term in
Equation \ref{ssimsol2}b they are associated with.  

In the ambipolar diffusion limit this solution reduces to equations 66--71 of
KK02, where $j=b_{\phi,s} \approx 0$. 

\bsp

\label{lastpage}

\end{document}